\documentclass[aps,prl,twocolumn,showpacs,groupedaddress]{revtex4}% review and submission
\usepackage{graphicx}  % needed for figures
\usepackage{dcolumn}   % needed for some tables
\usepackage{bm}        % for math
\usepackage{amssymb}   % for math
\usepackage{verbatim}

\usepackage{amsmath}
\usepackage{amsfonts}

\begin{document}

\title{Quasiparticle tunneling: P(E) theory}
\author{John M. Martinis$^1$}

\affiliation{$^{1}$Department of Physics, University of California,
Santa Barbara, California 93106, USA}

\date{\today}

\begin{abstract}
A calculation of the energy decay rate of a Josephson qubit from non-equilibrium quasiparticles is made using the environmental $P(E)$ theory.  For a large-capacitance qubit, we extend the theory to include the tunneling of quasiparticles, which has an electron- and hole-like charge components.

\end{abstract}

\maketitle

In this note, we calculate the dissipation due to the tunneling of quasiparticles using the environmental $P(E)$ theory\cite{refreview} to properly account for coupling to the qubit from the charge transfer.  We will first calculate the damping rate for a inductor-capacitor resonator using environmental theory; the prediction for phase qubits, for example, will be similar since qubit and harmonic oscillator matrix elements are nearly identical.  We will then consider the tunneling of quasiparticles, which has tunneling of charge carriers that is both electron- and hole-like.

Standard environmental theory gives a probability $P(E)$ for the environment to absorb energy from a tunneling event with charge transfer $q$
\begin{align}
P(E) &= \int_{-\infty}^{\infty} \frac{dt}{2\pi\hbar}
\exp[ J(t)+ \textrm{i}E t/\hbar ] \label{eqP}\\
J(t)&= \int_{-\infty}^{\infty} \frac{d\omega}{\omega}
\textrm{Re} \{ 2 Z(\omega)/R_K \} (e^{-\textrm{i}\omega t}-1) \label{eqJ}\ ,
\end{align}
where $Z(\omega)$ is the environmental impedance, $R_K=h/q^2$, and we have assumed for simplicity $T=0$.  We first consider an inductor in parallel with a Josephson junction of capacitance $C$, which has a resonance frequency $\omega_r=1/\sqrt{LC}$.  Here $L$ is the parallel combination of the external inductance and the Josephson inductance.  The impedance of the environment is
\begin{align}
Z(\omega) = \frac{\pi}{2C} [\delta(\omega+\omega_r)+\delta(\omega-\omega_r)]
\label{eqZ} \ ,
\end{align}
which describes how the resonator can both absorb or emit photons at frequency $\omega_r$.  Defining the resonator impedance $Z_r = 1/\omega_r C$ and inserting Eq.\,(\ref{eqZ}) into Eq.\,(\ref{eqJ}), we find
\begin{align}
P(E) &= \int_{-\infty}^{\infty} \frac{dt}{2\pi\hbar} e^{\textrm{i}E t/\hbar }
\exp[ \pi \frac{Z_r}{R_K} (e^{-\textrm{i}\omega_r t}-1)]  \label{eqP1} \ .
\end{align}
For the case of large capacitance resonators where $Z_r \ll R_K$, the second exponential can be expanded in a Taylor's series giving
\begin{align}
P(E) &\simeq (1-\pi Z_r / R_K)\delta(E) +
(\pi Z_r / R_K)\delta(E-\hbar\omega_r) \ .
\end{align}
The probability for the environment to absorb a photon is given by the second term, and can be rewritten as
\begin{align}
p(\hbar\omega_r)=\frac{q^2/2C}{\hbar\omega_r} \label{eqP0}\ .
\end{align}
This probability is much smaller than one, which makes physical sense since the charging energy $q^2/2C$ of the tunneling event is much smaller than the environmental energy $\hbar\omega_r$.  For the case where the environment has initially one photon, the probability $p(-\omega_r)$ for the environment to emit a photon during tunneling has the same magnitude as Eq.\,(\ref{eqP0}).

We have assumed in this calculation that charge $q=-e$ is transferred across the junction in a tunneling event.  This assumption must be modified since  quasiparticles are both electron ($q=-e$) and hole ($q=e$) like, and their interference introduces coherence factors to the total tunneling rate.  The environmental theory must be derived in a way that properly accounts both for the quasiparticle states and the transfer of charge to the environment.  The tunneling Hamiltonian is given by
\begin{align}
H_{T} =& \sum\limits_{L,R,\sigma}
 t_{LR}^{{}}   c_{L\sigma}^{\dag}c_{R\sigma}^{{}}   e^{-\textrm{i}\varphi}
+t_{LR}^{\ast }c_{L\sigma}^{{}}  c_{R\sigma}^{\dag }e^{+\textrm{i}\varphi}
\ ,
\label{HTterms}
\end{align}
where $t_{LR}^{{}}$ is the tunneling matrix element.  Changes to the internal (uncharged) state of the superconductor is described by the creation and annihilation operators.  The $L$ and $R$ indices describe states in the left and right superconductor, and $\sigma$ describes the two spin states of the electron.  The change in the electrical circuit to a tunneling event is represented by the charge displacement operators $e^{\pm\varphi}$, which describe the transfer of $\pm e$ from the tunneling event.  Here, $\varphi$ is the conjugate coordinate to charge $q$ and corresponds to a dimensionless flux with $d\varphi/dt = (e/h) V = (e/h) (q/C)$.  It has a commutation relation $[q,\varphi]=\textrm{i} e$ and gives a charge displacement operator according to the relation $e^{\textrm{i}\varphi}qe^{-\textrm{i}\varphi}=q-e$.

The electron operators need to be re-expressed in terms of the quasiparticle operators $\gamma $ because quasiparticles are eigenstates of the superconducting leads.  The four electron operators are given by
\begin{equation}
\begin{array}{cc}
c_{k\uparrow}=
u_{k}\gamma _{k0\uparrow}+v_{k}\gamma_{k1\downarrow}^{\dagger }
& c_{k\downarrow}=
u_{k}\gamma _{k0\downarrow}-v_{k}\gamma _{k1\uparrow}^{\dagger } \\
c_{k\uparrow}^{\dagger }=
u_{k}\gamma _{k0\uparrow}^{\dagger }+v_{k}\gamma _{k1\downarrow}
& \ \ c_{k\downarrow}^{\dagger }=
u_{k}\gamma_{k0\downarrow}^{\dagger }-v_{k}\gamma _{k1\uparrow}
\ .
\end{array}
\label{ceqs}
\end{equation}
The standard electron and hole occupation factors  $u_k$ and $v_k$ are given by
\begin{align}
u_k^2&=(1/2)(1+\xi_k/E_k) \label{equ} \\
v_k^2&=(1/2)(1-\xi_k/E_k) \label{eqv} \ ,
\end{align}
where $\xi_k$ is the energy of the electron/hole states referred to the Fermi level, $E_k=\sqrt{\xi_k^2+\Delta^2}$ is the quasiparticle energy, and $\Delta$ is the superconducting gap.

The tunneling Hamiltonian, written in terms of the quasiparticle operators, is found by substituting Eq.\,(\ref{ceqs}) into Eq.\,(\ref{HTterms}).  Eight of the resulting terms correspond to tunneling of a quasiparticle across the junction.  For example, the tunneling from the state $L0$ to $R0$ is given by
\begin{align}
\overrightarrow{H}_{T}= t_{LR} ( u_L^{{}} u_R^{{}}e^{-\textrm{i}\varphi} - v_L^{{}} v_R^{{}} e^{\textrm{i}\varphi} ) \gamma_{L0}^{{}} \gamma_{R0}^\dagger \ ,
\end{align}
where we have used $t_{LR}=t_{LR}^\ast$. As seen from the charge displacement operators, the two terms corresponds to charge tunneling in opposite directions.

The tunneling of a quasiparticle produces both an electron-like and hole-like transfer of charge to the environment.  The superposition of charge transfer has not been considered previously for environmental theory, and can be incorporated readily into the formalism with only a change in the environmental displacement correlator (see Eq.\,(44) in Ref.\,\cite{refreview})
\begin{align}
 e^{J(t)} & = \langle e^{\textrm{i}\varphi(t)} e^{- \textrm{i}\varphi(0)} \rangle  \\
  & \rightarrow  \langle
( u e^{-\textrm{i}\varphi(t)} - v e^{\textrm{i}\varphi(t)} )
( u e^{\textrm{i}\varphi(0)} - v e^{-\textrm{i}\varphi(0)} )
\rangle \label{eqcorr}
\ ,
\end{align}
where $u=u_L^{{}} u_R^{{}}$ and $v=v_L^{{}} v_R^{{}} $.  The 4 correlators can be simplified by noting that for a harmonic oscillator the function is Gaussian, and therefore determined by its first and second moments \cite{refreview}.  Expanding terms to second order, one finds
\begin{align}
& \langle e^{\pm\textrm{i}\varphi(t)} e^{\mp \textrm{i}\varphi(0)} \rangle \nonumber \\
&\simeq
\langle [1\pm \textrm{i}\varphi(t)+(\textrm{i}\varphi(t))^2/2]
        [1\mp \textrm{i}\varphi(0)+(\textrm{i}\varphi(0))^2/2] \\
&=\langle1\pm\textrm{i}[\varphi(t)-\varphi(0)]+\varphi(t)\varphi(0)
-[\varphi(t)^2+\varphi(0)^2]/2 \rangle \nonumber \\
&=\langle1+[\varphi(t)-\varphi(0)]\varphi(0)\rangle \\
&\simeq  e^{\langle [\varphi(t)-\varphi(0)]\varphi(0)\rangle} \ ,
\end{align}
where we have used $\langle \varphi(t)-\varphi(0) \rangle = 0$ and
$\langle \varphi(t)^2 \rangle = \langle \varphi(0)^2 \rangle$.  One similarly finds
\begin{align}
 \langle e^{\textrm{i}\varphi(t)} e^{ \textrm{i}\varphi(0)} & +
e^{-\textrm{i}\varphi(t)} e^{- \textrm{i}\varphi(0)}\rangle \\
&\simeq
\langle 2 - 2\varphi(t)\varphi(0)
-[\varphi(t)^2+\varphi(0)^2] \rangle \\
&= 2\langle1-[\varphi(t)+\varphi(0)]\varphi(0)\rangle \\
&\simeq 2 e^{-\langle [\varphi(t)+\varphi(0)]\varphi(0)\rangle} \ .
\end{align}
Equation (\ref{eqcorr}) can thus be written as
\begin{align}
e^{J(t)} = &[(u^2+v^2) e^{\langle\varphi(t)\varphi(0)\rangle}
-2uv e^{-\langle \varphi(t)\varphi(0)\rangle} ] \nonumber \\
& \times e^{-\langle\varphi(0)\varphi(0)\rangle} \label{eqJnew}\ .
\end{align}
Note that the quantity $\langle \varphi(t)\varphi(0)\rangle$ is the noise correlation function for the (dimensionless) flux fluctuations.  As shown in Eq. (\ref{eqJ}), it can be computed as the Fourier transform of the noise spectrum obtained from the fluctuation dissipation theorem.

For the case $Z_r \ll R_K$, corresponding to large capacitance junctions where $\varphi$ is small, the exponential in Eq. (\ref{eqJnew}) can be expanded in a Taylor series
\begin{align}
e^{J(t)} \simeq (u-v)^2 + (u+v)^2\langle\varphi(t)\varphi(0)\rangle \label{eqJexp}\ .
\end{align}
Because the density of quasiparticles are small and they mostly thermalize to energies close to the gap, the occupation probabilities are all approximately the same $u_L^2=v_L^2=u_R^2=v_R^2 =1/2$. In this limit, the probability for a quasiparticle to tunnel without a change in energy is given by the $(u-v)^2 \simeq 0$, showing a large suppression of tunneling due to the electron- and hole-like nature of the quasiparticle.  However, the coherence factors when absorbing or emitting a photon to the environment is proportional to the second term, which is near-unity  $u^2+v^2 \simeq 1$.

The coherence factors can be computed using Eqs.\,(\ref{equ}) and (\ref{eqv})
\begin{align}
(u_L^{{}}u_R^{{}} \pm v_L^{{}}v_R^{{}})^2 &=
\frac{1}{2} \frac{E_LE_R+\xi_L\xi_R \pm \Delta_L\Delta_R}{E_LE_R} \\
&= \frac{1}{2} \frac{E_LE_R\pm \Delta_L\Delta_R}{E_LE_R} \ ,
\end{align}
where in the second equation the terms $\xi_L\xi_R$ have been averaged to zero for typical integrations over quasiparticle states.

We first compute the quasiparticle tunneling rate (from left to right) corresponding to the first term in Eq.\,(\ref{eqJexp}).  This process is equivalent to a thermal current across the junction, as there is no charge transfer from tunneling.  The average tunneling rate is
\begin{align}
\overrightarrow{\Gamma}_T= &  \frac{1}{e^2 R_t} \int_\Delta^\infty dE\,2\rho_L 2 \rho_R
(u-v)^2  f_L(E)[1-f_R(E)] \\
= & \frac{2}{e^2R_t}\int_\Delta^\infty dE \,
\frac{E^2-\Delta^2}
{(\sqrt{E^2-\Delta^2}\,)^2}  f_L(E)[1-f_R(E)] \\
= & \frac{2}{e^2R_t}\int_\Delta^\infty dE \,  f_L(E)[1-f_R(E)]
\label{eqGammaT} \ ,
\end{align}
where $R_T$ is the (normal state) resistance of the tunnel junction, and $\rho=E/\xi=E/\sqrt{E^2-\Delta^2}$ is the normalized quasiparticle density of states.  The factor of 2 for each $\rho$ accounts for summing over both positive and negative $\xi$.  Note that without interaction to the environment, the final energy of the quasiparticle excitation is unchanged from the initial energy.  The integral, having only occupation factors, corresponds to simple tunneling of the excitations through the junction, as appropriate for quasiparticles diffusing through the superconductor.

The second term in Eq.\,(\ref{eqJexp}) corresponds to a tunneling event that couples energy from the qubit to a quasiparticle.  The rate for this process has been written down in the main article.

\section{QP damping in a coplanar resonator}

In this section we compute the Q-factor (damping) for a coplanar resonator due to non-equilibrium quasiparticles in the superconductor.  The discussion follows the Ph.D. thesis of J. Gao \cite{Gao}.

A bulk superconductor has complex conductivity $\sigma=\sigma_1-i\sigma_2$ given by the Mattis-Bardeen theory
\begin{align}
\frac{\sigma_1}{\sigma_n} &\simeq \sqrt{2} \Big( \frac{\Delta}{\hbar \omega} \Big) ^{3/2} \frac{n_\textrm{qp}}{D(E_F) \Delta } \ ,\\
\frac{\sigma_2}{\sigma_n} &\simeq \pi \frac{\Delta}{\hbar\omega} - 2\sqrt{2} \Big( \frac{\Delta}{\hbar \omega} \Big) ^{3/2} \frac{n_\textrm{qp}}{D(E_F) \Delta }\ ,\label{eqs2}\\
&= \pi \frac{\Delta}{\hbar\omega} -2\frac{\sigma_1}{\sigma_n}\ ,\label{eqs22}
\end{align}
where parameters have been defined in the main text, and $\sigma_n$ is the normal state conductivity.  We will assume that the quasiparticle density is small, giving $\sigma_1 \ll \sigma_2$ and little dissipation.

The surface impedance of a superconductor depends on details of the superconductor and interface.  In the thin film limit, the surface impedance is obtained via the geometrical factor of the film thickness $d$
\begin{align}
Z_s =\frac{1}{d}\frac{1}{(\sigma_1-i\sigma_2)}
 \simeq \frac{1}{d}\Big( \frac{\sigma_1}{\sigma_2^2}+i\frac{1}{\sigma_2} \Big) \ .\label{eqzsthin}
\end{align}
The ratio of the real to imaginary impedance is
\begin{align}
\frac{\textrm{Re}(Z_s)}{\textrm{Im}(Z_s)}  = \gamma\frac{\sigma_1}{\sigma_2} \ , \label{defgamma}
\end{align}
where here, for the thin film limit, $\gamma = 1$.  For the case of thick films in the local (dirty) limit, the surface impedance is given by standard electromagnetic theory
\begin{align}
Z_s =\Big( \frac{i\mu_0\omega}{\sigma_1-i\sigma_2}\Big)^{1/2}\ . \label{eqzthick}
\end{align}
For $\sigma_1 \ll \sigma_2$, the ratio of the impedances is the same form as Eq.\,(\ref{defgamma}), but with $\gamma = 1/2$ in the local (dirty) limit and 1/3 in the extreme anomalous (clean) limit \cite{Gao2.80}.

The quality factor $Q$ may be computed taking into account the field distribution in the coplanar transmission line.  For a resonance frequency $\omega$, standard microwave theory gives $1/Q=\mathcal{R}/\omega \mathcal{L}$, the ratio of the dissipative to total dispersive impedance.  Here, $\mathcal{L}=\mathcal{L}_m+\mathcal{L}_s$ and $\mathcal{R}=(g/s)\textrm{Re}(Z_s)$ are the total inductance and resistance per length of the line, $\mathcal{L}_m$ is the geometric inductance coming from the magnetic field, $\mathcal{L}_s=(g/s)\textrm{Im}(Z_s/\omega)$ is the kinetic inductance, $s$ is the width of the center line, and $g \simeq 1.0 - 1.2$ is a geometric factor accounting for the coplanar geometry \cite{Gao3.1.3}.  Defining the fractional contribution of the kinetic inductance as $\alpha = \mathcal{L}_s/\mathcal{L}$, one finds
\begin{align}
\frac{1}{Q} &= \alpha \gamma \frac{\sigma_1}{\sigma_2} \ .\label{Qeq}
\end{align}
For typical transmission lines the effect of the kinetic inductance is small, corresponding to $\alpha \ll 1$.

In the thin-film limit, the fractional kinetic inductance using Eq.\,(\ref{eqzsthin})  is $\alpha \simeq g/(d s \sigma_2)/\mathcal{L}_m$, which gives
\begin{align}
\frac{1}{Q} = \Big[\frac{\mathcal{R}_n}{\omega\mathcal{L}_m}\Big]\, g
\frac{\sqrt{2}}{\pi^2} \Big( \frac{\hbar\omega}{\Delta} \Big) ^{1/2} \frac{n_\textrm{qp}}{D(E_F) \Delta } \ ,
\end{align}
where $\mathcal{R}_n = 1/ds\sigma_n$ is the normal state resistance per unit length of the coplanar line.  The first term is the expected damping factor for a normal-state transmission line, while the remaining factors represents the suppression of dissipation due to the small number of quasiparticles in the superconducting state.

In the thick-film limit, the inductances may be parameterized by $\textrm{Im}(Z_s/\omega)\equiv\mu_0\lambda$ and $\mathcal{L}_m\equiv\mu_0 g_m$.  In the local (dirty) limit, the penetration depth may be obtained from Eq.\,(\ref{eqzthick})
\begin{align}
\lambda = \Big(\frac{1}{\mu_0\omega\sigma_2}\Big)^{1/2}
= \Big(\frac{1}{\pi\mu_0\sigma_n\Delta/\hbar}\Big)^{1/2} \ ,
\end{align}
with a magnitude $\lambda \sim 50\,\textrm{nm}$ for Al \cite{Gao2.2.6}.  The magnetic inductance is described by a geometrical factor $g_m \simeq 0.31$ that is only logarithmicly dependent on design parameters \cite{Gao3.1.3}.  Using Eq.\,(\ref{Qeq}), the quality factor is given by
\begin{align}
\frac{1}{Q} = \frac{\lambda}{s}\  \frac{g}{g_m}
\gamma \frac{\sqrt{2}}{\pi} \Big( \frac{\Delta}{\hbar\omega} \Big) ^{1/2} \frac{n_\textrm{qp}}{D(E_F) \Delta } \ .
\end{align}
Note that the $Q$ scales as $\sigma_n^{1/2}$, which is a smaller power law than for the thin-film limit.

We have not yet accounted for the change $\delta\sigma_2$ from $n_\textrm{qp}$ in Eqs.\,(\ref{eqs2}) and (\ref{eqs22}), which slightly decreases the kinetic inductance and increases the resonance frequency.  The fractional change in frequency is computed as $\delta\omega/\omega=-\,\delta \mathcal{L}_s/2\mathcal{L}=-\alpha\,\delta\mathcal{L}_s/2\mathcal{L}_s
=-\alpha\gamma\,\delta\sigma_2/2\sigma_2 = \alpha\gamma\,\delta\sigma_1/\sigma_2 = 1/Q$,
so that the fractional frequency change is the same magnitude as $1/Q$.

\end{document}